\documentclass[fleqn,usenatbib]{mnras}

\usepackage{newtxtext,newtxmath}
\usepackage[T1]{fontenc}
\usepackage{ae,aecompl}

\DeclareRobustCommand{\VAN}[3]{#2}
\let\VANthebibliography\thebibliography
\def\thebibliography{\DeclareRobustCommand{\VAN}[3]{##3}\VANthebibliography}

\usepackage{graphicx}	
\usepackage{amsmath}	

\title[Internal structure of clusters]{The VISCACHA survey – VI. Dimensional study of the structure of 82 star clusters in the Magellanic Clouds}
\author[Rodriguez et al.]{
M. Jimena Rodríguez$^{1,2,3}$\thanks{E-mail: jimenaro@fcaglp.unlp.edu.ar},
C. Feinstein$^{2,3,4}$,
G. Baume$^{2,3,4}$,
B. Dias$^{5}$, 
F. S. M. Maia$^{6}$,
J. F. C. Santos Jr.$^{7}$,
\newauthor
L. Kerber$^{8}$,
D. Minniti$^{9,10}$,
A. P\'erez-Villegas$^{11}$,
B. De Bórtoli$^{2,3,4}$,
M. C. Parisi$^{3,12,13}$ \&
R. A. P. Oliveira$^{14}$
\\
$^{1}$Steward Observatory, University of Arizona, 933 N Cherry Ave,Tucson, AZ 85721, USA \\
$^{2}$Instituto de Astrofísica de La Plata, CONICET--UNLP, Paseo del Bosque S/N, B1900FWA La Plata, Argentina \\
$^{3}$ Consejo Nacional de Investigaciones Científicas y T\'ecnicas, Godoy Cruz 2290, C1425FQB, Ciudad Aut\'onoma de Buenos Aires, Argentina\\
$^{4}$Facultad de Ciencias Astron\'omicas y Geof\'isicas, UNLP, Paseo del Bosque S/N, B1900FWA La Plata, Argentina \\
$^{5}$Instituto de Alta Investigaci\'on, Universidad de Tarapac\'a, Av. Luis Emilio Recabarren 2477, Iquique, Chile \\
$^{6}$Instituto de F\'isica - Universidade Federal do Rio de Janeiro, Av. Athos da Silveira Ramos, 149, Rio de Janeiro, 21941-909, Brazil \\
$^{7}$Departamento de F\'isica, ICEx - UFMG, Av. Ant\^onio Carlos 6627, Belo Horizonte 31270-901, Brazil\\
$^{8}$Departamento de Ciências Exatas e Tecnológicas, UESC, Rodovia Jorge Amado km 16, 45662-900, Brazil \\
$^{9}$Instituto de Astrof\'isica, Facultad de Ciencias Exactas, Universidad Andres Bello, Av. Fern\'andez Concha 700, Santiago, Chile\\
$^{10}$ Vatican Observatory, V00120 Vatican City State, Italy\\
$^{11}$Instituto de Astronom\'ia, Universidad Nacional Aut\'onoma de M\'exico, Apartado Postal 106, C. P. 22800, Ensenada, B. C., Mexico\\
$^{12}$Observatorio Astron\'omico, Universidad Nacional de C\'ordoba, Laprida 854, X5000BGR, C\'ordoba, Argentina.\\
$^{13}$Instituto de Astronom{\'\i}a Te\'orica y Experimental (CONICET-UNC), Laprida 854, X5000BGR, C\'ordoba, Argentina\\
$^{14}$Universidade de S\~ao Paulo, IAG, Rua do Mat\~ao 1226, Cidade Universit\'aria, S\~ao Paulo 05508-900, Brazil
}

\date{Accepted XXX. Received YYY; in original form ZZZ}

\pubyear{2022}

\begin{document}
\label{firstpage}
\pagerange{\pageref{firstpage}--\pageref{lastpage}}
\maketitle
\begin{abstract}
We present a study of the internal structure of 82 star clusters located at the outer regions of the Large Magellanic Cloud and the Small Magellanic Cloud  using data of the VISCACHA Survey. Through the construction of the {\it minimum spanning tree}, which analyzes the relative position of stars within a given cluster, it was possible to characterize the internal  structure and explore the fractal or subclustered distribution for each cluster. We computed the parameters  $\overline{m}$ (which is the average length of the connected segments normalized by the area), $\overline{s}$  (which is the mean points separation in units of cluster radius), and $Q$ (the ratio of these components).  These parameters are useful to  distinguish between radial, homogeneous, and substructured distributions of stars. The dependence of these parameters with the different characteristics of the clusters, such as their ages and spatial distribution, was also studied.
We found that most of the studied clusters present a homogeneous stellar distribution or a distribution with a radial concentration. Our results are consistent with the models, suggesting that more dynamically evolved clusters seem to have larger $Q$ values, confirming previous results from numerical simulations. There also seems to be a correlation between the internal structure of the clusters and their galactocentric distances, in the sense that for both galaxies, the more distant clusters have larger $Q$ values. 
We also paid particular attention to the effects of contamination by non-member field stars and its consequences finding that field star decontamination is crucial for these kinds of studies.


\end{abstract}




\section{Introduction}


Star clusters form in dense clumps inside giant molecular clouds (GMCs). The formation of these clumps is a combined action of supersonic turbulence and self-gravity causing a hierarchical fragmentation of the cloud. These hierarchical collapse will continue inside the clumps, and stars will be formed in the densest part. These processes lead to a fractal or sub-clustered initial distribution of stars inside the cluster \citep{2010RSPTA.368..733C}. Numerical and observational results suggest that this initial fractal distribution is erased with time due to the dynamic evolution of the cluster \citep{2004MNRAS.348..589C}. In this process unbound clusters rapidly dissolve into the field, whereas bound clusters tend to a central concentration of stars with a smooth radial density profile. However this is only a general picture. We barely know about the time scale in which star clusters loose their initial fractal distribution. The dynamic evolution of clusters is a complex process that depends of several not yet well understood factors \citep{2014MNRAS.438..620P}. In fact, there exist clusters as young as 1~Myr with a central concentration of stars \citep[e.g. $\rho$ Ophiuchus,][]{2004MNRAS.348..589C}{} and older clusters that present a substructure distribution. For example, \cite{2009ApJ...696.2086S} found substructures in NGC~1513 and NGC~164 older than 100~Myr. On the other hand, recent work of \cite{2020MNRAS.493.4925D} shows by dynamical modelling that there is some evolution towards a centrally concentrated distributions of stars as early as 1~Myr in same cases.

The study of a relatively large and homogeneous sample of star clusters covering a large range of ages can help searching for an age limit beyond which there is no signature of an internal fractal structure. The sample of 82 SMC and LMC clusters analyzed by \citet{2020MNRAS.498..205S} span ages from $\sim 30$ Myr to $5.5$ Gyr, and is part of the "VIsible Soar photometry of star Clusters in tApii and Coxi HuguA" (VISCACHA) Survey \citep{2019MNRAS.484.5702M}. \citet{2020MNRAS.493.4925D} analysed the global structure of these clusters, whereas we make a deeper analysis on the fluctuations on the internal structure of these clusters in the present work. In this context, \cite{2020MNRAS.498..205S} investigated the stellar density and surface brightness profiles of star clusters on the Magellanic Clouds. They obtained structural parameters by fitting King´s \citep{1962AJ.....67..471K} and Elson´s models \citep{1987ApJ...323...54E} to the observed profiles. In the present study, the same data were analyzed from a new perspective, complementing the work of \cite{2020MNRAS.498..205S}.

On the other hand, the time that a star cluster require to dynamically erase any signature of their initial fractal structure depends not only on the chronological time, but also on the mass and size of the cluster \citep{2011MNRAS.410L...6G}. This information is available from \citet{2020MNRAS.498..205S} and therefore it is now possible to make a self-consistent analysis of how the fractal structure evolves using the chronological and dynamical age as baselines.
 
The paper is organized as follows. In Section \ref{sec:data} we describe the data and in Section \ref{sec:method} we present the applied methodology. Section \ref{sec:results} details the main characteristics obtained for the studied clusters. Finally, in Section  \ref{sec:discussion} we present a general discussion and Section \ref{sec:conclusions}, we draw our conclusions.

\begin{figure*}
	\centering
	\includegraphics[width=0.8\textwidth]{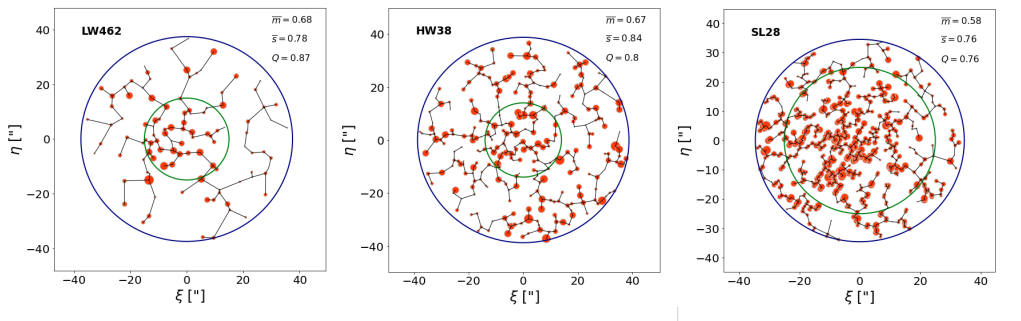}
	\caption{MST for three of the studied clusters. The cluster in the left present a radial concentration of stars, the cluster in the middle panel present an homogeneous distribution, and the cluster in the right present a fractal distribution. The green and blue circles correspond to $r_c$ and $r_h$ respectively. The size of the points indicate $V$ magnitude of the stars. $\xi$, $\eta$ coordinates are relative to the corresponding clusters centers.}
	\label{fig:MST}
\end{figure*}

\section{Data} 
\label{sec:data}

We use  data from the VISCACHA Survey,  consisting of photometric observations with $BVI$ filters obtained with the 4.1~$m$ SOAR telescope using the SOAR Adaptive Module (SAM, \citealt{2016PASP..128l5003T}). The first data release covers the fields of 83 clusters located in the outermost regions of the Large Magellanic Cloud (LMC) and the Small Magellanic Cloud (SMC), see \cite{2019MNRAS.484.5702M} and \citet{2020IAUS..351...89D} for details. 

In particular, we studied 51 clusters located in the LMC with projected distances between 4.5 and 6.5 kpc to the center of this galaxy and 31 clusters in the surroundings of the SMC with projected distances to the center between 1 and 6.5 kpc. The selected clusters span a wide age range between 40 Myr and 5.5 Gyr and concentration parameters $c = log(r_{\rm tidal}/r_{\rm core})$ between $\sim 0.4$ and 1.9. \citep{2020MNRAS.498..205S}.

\section{Methodology} 
\label{sec:method}

 The study of the structure of stellar systems can be carried out from a statistical point of view using different distribution functions. In this case, the ``Distance Distribution Function'' \citep{1995MNRAS.272..213L}, the ``Correlation Function'' or an alternative of the latter known as the ``Two-Point Correlation Function'' (TPCF), 
could be used. In particular, the first one is associated with the ``Mean normalized correlation length'' ($\bar{s}$). Although these functions generally allow a description of the analyzed structures in some detail, in some cases they fail to clearly differentiate whether it is a fractal structure. A suitable complement for the analysis is provided by the use of a ``minimum spanning tree'' (MST, \citealt[][]{10.2307/2346439}) of the 2D projection of the studied stellar structure. In this case, the associated parameter is the ``Mean normalized edge length'' ($\bar{m}$). Fig.\ref{fig:MST} is an example of the MST plotted over the sky projection of three clusters.

\subsection{The minimum spanning tree and Q parameter}
\label{Sec:MST}

To study the internal structure of the star clusters, we use the MST method. The MST of a distribution of points is the shortest network of straight lines connecting all the points without closed loops. From the MST we estimated the parameters $\overline{m}$,  and $\overline{s}$. These values are given by the following equations:

$\displaystyle\overline{m}=\frac{1}{(AN)^{1/2}}$$\displaystyle\sum_{i=1}^{N-1}m_i$

\noindent where $N$ is the number of stars, $A$ is the area of the smallest circle that contains all the stars, and $m_i$ is the length of the $i$ segment of the MST. So, $\overline{m}$ is the average length of the connected segments normalized by the cluster area,

$\displaystyle\bar{s}=\frac{2}{N(N-1)R_{sc}}\displaystyle\sum_{i=1}^{N-1}\displaystyle\sum_{j=i+1}^{N}\left|\vec{r_i}-\vec{r_j} \right|$,

\noindent where $r_i$ is the position of the $i$ object while $r_j$ is the position of another object named  $j$. $\overline{s}$ is the mean separation of the points normalized by the cluster radius. 
From the MST we derived the $Q$ parameter defined as 
$Q$~=~$\overline{m}$/$\overline{s}$. 
This parameter was first introduced by \cite{2004MNRAS.348..589C} as a tool to study the distribution of stars in a cluster. They worked with simulated clusters using both radial density distribution of stars, which follows the form $r^{-\alpha}$ where $\alpha$ is the concentration factor, and fractal distributions with different fractal dimensions. They concluded that clusters with a substructure distribution, which is usually compared with a fractal distribution present values of $Q < 0.8$. In uniform stellar distributions, they observed that $Q$ was $\sim 0.8 $. In contrast, values of $ Q> 0.8 $ corresponded to radial distributions with a central concentration of stars. Since then, the MST and $Q$ parameter have been widely used to classify the internal structure of stars clusters \citep[e.g.][]{2015MNRAS.448.2504G, 2019MNRAS.490.2521H,2020A&A...644A.101R}.
We estimated errors in $\overline{m}$, $\overline{s}$, and $Q$ using the bootstrap method \citep[e.g.][]{efron1992bootstrap}.

\cite{2019MNRAS.490.2521H} studied the uncertainties in $Q$ as a function of the number of members. They estimated the values of $Q$, $\overline{s}$, and $\overline{m}$ for two clusters with more than 100 stars. Then they decreased the number of stars and calculated how these parameters change. They found that these values were almost constant for $N>70$ stars and a lower limit of 20 stars before the values of parameters diverge. The clusters studied in our sample, present between 14 and 599 stars inside the half-light radius ($r_{h}$). While $\sim 50 \%$ of the clusters have less than 70 stars, only 4 clusters ($\sim 5 \%$) have less than 20 stars. So it is expected that the values obtained in this work do not have a significant error.

\subsection{Contamination by non-member stars}
\label{contamination}

The estimated $Q$ values could be affected by the inclusion of non-member field stars. We tried to minimize this contamination using only brighter stars with lower magnitude errors, within a limited radius around the cluster center (see Sect.~\ref{sec:results}), but some non-member stars could still be present in our computations.

To calculate how significant is the effect of the contaminants, we selected 15 clusters from our sample spanning a wide parameter range since the membership probability of the stars in these cases was determined by \citet{2019MNRAS.484.5702M,2021A&A...647L...9D, 2022MNRAS.512.4334D} using the procedure described in \citet{2010MNRAS.407.1875M}, briefly presented here. 
The method is based on the comparison between the CMD of stars within the cluster radius with a CMD of stars located in a nearby field around the cluster area. Each box of colour and magnitude in the grid of the cluster CMD is compared with the field CMD, and if the cluster stars outnumber the field stars, then they are more likely cluster members. This procedure is repeated for all boxes, and for different grid sizes and positions. At the end, each star will have a series of different probabilities,
The final probability for a given star is the median of all assigned probabilities. Here, we test only assigned members (membership > 0) and high probable members (membership > 0.5). This decontamination procedure has been applied in the previous VISCACHA papers. For more details, see \cite{2010MNRAS.407.1875M}.

\begin{figure}
	\centering
	\includegraphics[width=0.95\columnwidth]{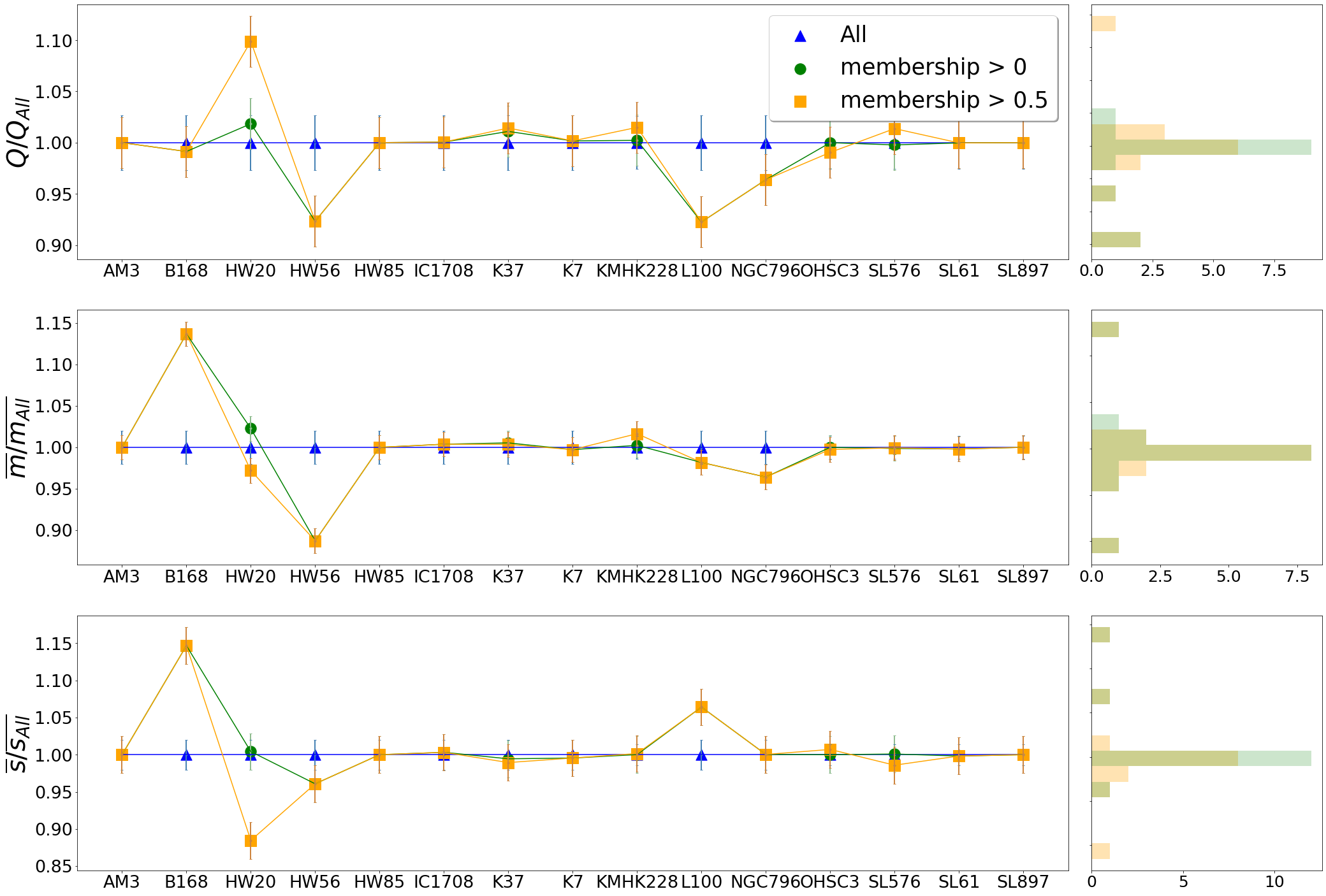}
	\caption{Values of $Q, \overline{m}, \overline{s}$ estimated without considering membership (blue), taking into account the assigned members (green), and the high probable members (yellow). The values are normalized to those obtained without considering membership. The errors were estimated using the bootstrap method. The histograms in the right show the relative differences in the estimated parameters considering all the
    assigned members (green), and only the high probable members (orange).}
	\label{fig:delta_Q_memb}
\end{figure}

\begin{figure}
	\centering
	\includegraphics[width=0.95\columnwidth]{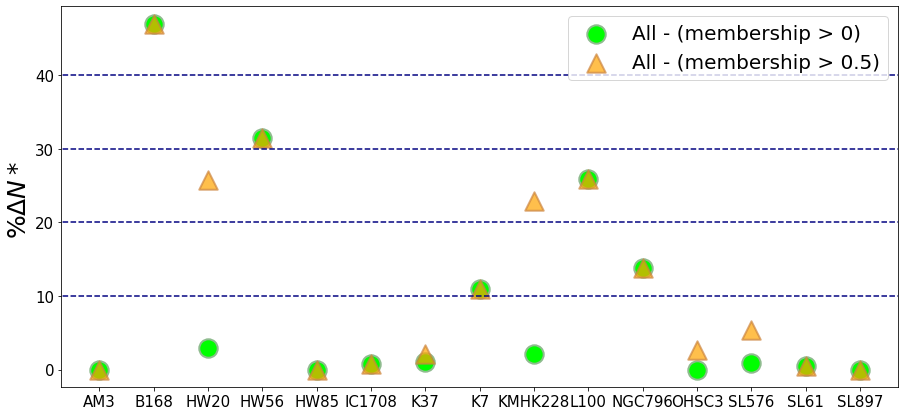}
	\caption{Level of contamination by field stars within the cluster radius. The relative difference between the total number of stars and those considered members is shown for a threshold in membership $> 0$ and $>0.5$ as green circles and orange triangles, respectively.}
		\label{fig:delta_N_memb}
\end{figure}

We estimated the $\overline{m}$, $\overline{s}$ and $Q$ values for this sample using all assigned members and high probable members. 
In Fig.~\ref{fig:delta_Q_memb} we compared the obtained values with those that were estimated without considering membership (see Sect.~\ref{sec:results}). We found that the difference in these quantities are very low, and only in one cluster the relative values of $Q$ and $\overline{m}$ reaches 1.15. 
The cluster with the largest variation in $Q$ is HW20, followed by L100 and HW56. HW20 and HW56 have low number of stars ($\lesssim$ 50) and the variation of the number of stars using membership and without considering it, represent a considerable percentage of the total number of stars ($\sim30\%$), as can be seen in Fig.~\ref{fig:delta_N_memb}. The cluster L100 presents a considerable number of stars ($\sim 100$), but the difference of stars using membership and without consider it is also an important value ($\sim 25\%$), explaining the observed variation in $Q$.
In Fig.~\ref{fig:delta_N_memb} we also can note that for B168 the difference of stars is near $50\%$. Even though this cluster do not present variation in $Q$ (Fig.~\ref{fig:delta_Q_memb}), it has highest variation in the $\overline{s}$ and $\overline{m}$ parameters.

\begin{figure}
	\centering
	\includegraphics[trim=0 30 0 0, width=\columnwidth]{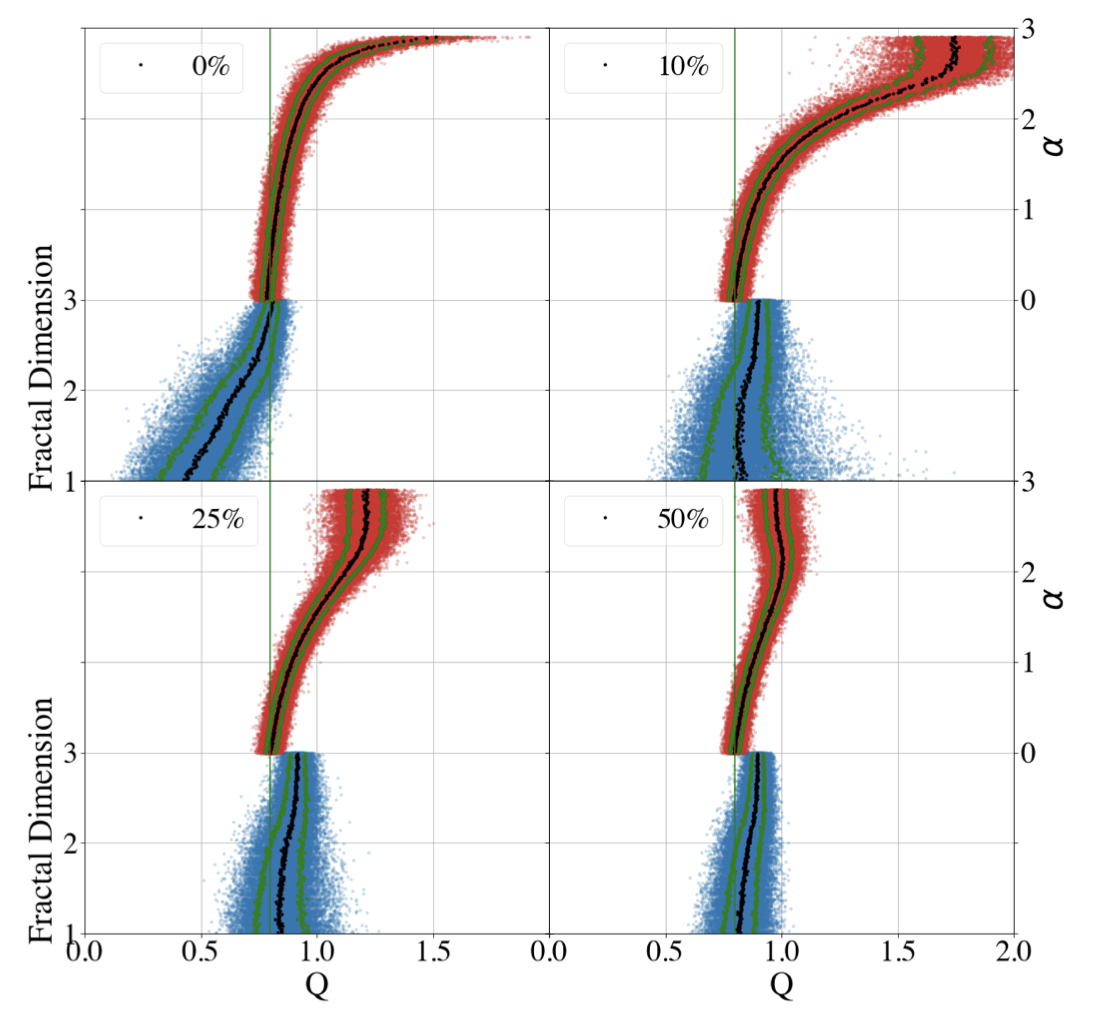}
	\caption{Simulations of clusters with different contamination levels with a set of non-members stars (0\%,10\%,25\% and 50\%). In each panel the blue points are clusters of fractal dimension vs the measure Q. The red dots are non-fractal clusters with a different parameter $\alpha$ that accounts for the exponential distributions of stars in the cluster vs the measure Q. The black and green lines represent the mean value and the standard deviation.}
	\label{fig:conta}
\end{figure}

\begin{figure}
	\centering
	\includegraphics[trim=0 30 0 0, width=0.95\columnwidth]{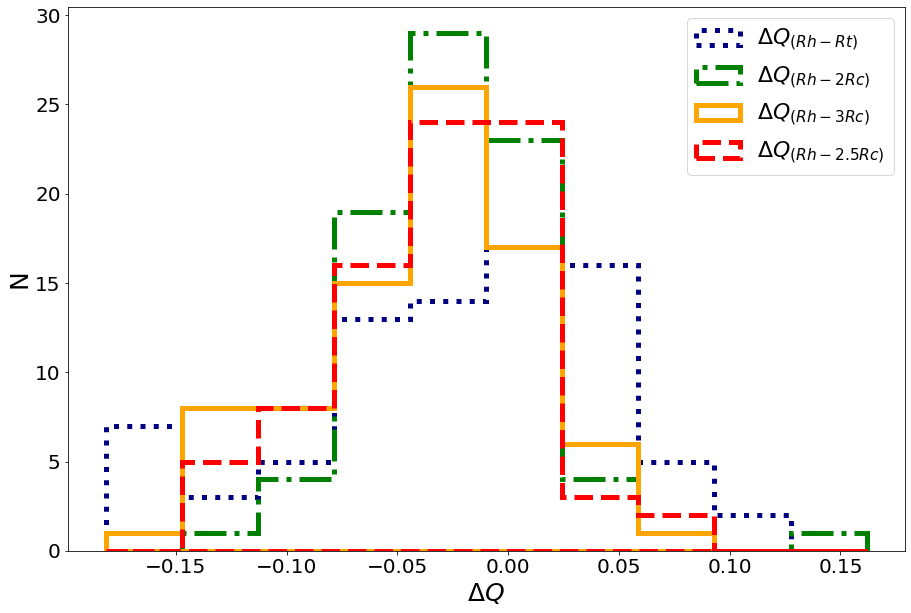}
	\caption{Variations in the obtained $Q$ value using different radius.}
	\label{fig:delta_Q_rad}
\end{figure}

\begin{figure*}
	\centering
	\includegraphics[width=0.8\textwidth]{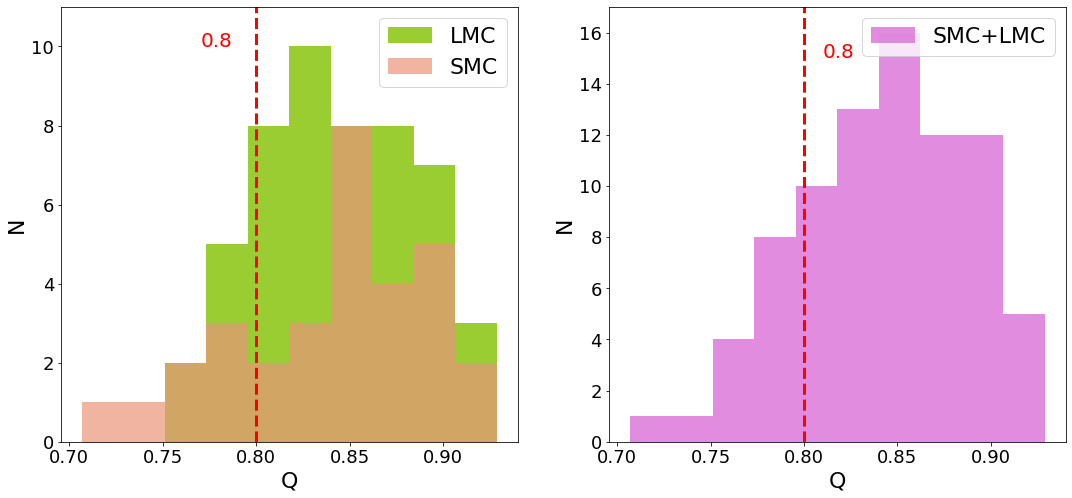}
	\caption{\emph{Left:} Histogram of the $Q$ values for the LMC (green) and the SMC (orange).
		\emph{Right:} Total histogram of the $Q$ values. The vertical red dashed line indicate the limit value of $Q=0.8$. }
	\label{fig:histQ}
\end{figure*}

We also, made  simulations of clusters with fractal and radial exponential distribution of stars (non fractal) to further analyze the effect of non-member stars in the final value of Q. First, we contaminated them with a uniform distribution of stars in order to simulated a constant density of background or foreground field stars and measured the new contaminated $Q$ parameter. We could then obtain on average, how the different levels of contamination could mislead our observed measure of $Q$ on the real clusters. To make the mock clusters, we  follow  the recipe given by \cite{2004MNRAS.348..589C} for the fractal and non-fractal clusters.
To evaluate our results (Fig.~\ref{fig:conta}), we follow the strategy by \cite{2004MNRAS.348..589C}. This figure is a combination of two plots: a) the bottom one is the fractal dimension used to generate the cluster vs the measured $Q$, where the fractal dimension increases with decreasing fractal structures, while  b) the upper an the exponential parameter ($\alpha$) as is used in \cite{2004MNRAS.348..589C}, which accounts for the central concentration of the radial star density for a non-fractal cluster vs the measured $Q$. 
Each panel of Fig.~\ref{fig:conta} corresponds to a different degree of contamination from 0\% to 50\%.  

From Fig.~\ref{fig:conta} it is clear that for the fractal clusters, only a 10\% of background stars could already hide the fractal structure, as all $Q$ values move from $Q < 0.8$ to $Q > 0.8.$. But in the case of the evolved cusp clusters  (the top region of the plots), the effect of the contamination on the Q values is almost negligible. Therefore, these clusters still have a $Q > 0.8$ and they maintain their status as non-fractal. 

The 82 clusters analyzed in this work present measured $Q$ without the membership information within $0.7 \lesssim Q \lesssim 0.9$. Looking at Fig.~\ref{fig:conta} and assuming that these 82 are non-fractal, their $Q$ values should not change much regardless the level of contamination. Maybe the largest $Q\approx0.9$ in the sample could mean a non-fractal cluster with an original $Q\approx0.8$ and about 50\% of contamination. However, if all 82 clusters were fractal, all the $Q$ value would be consistent with a 25-50\% of contamination level for an original range of $0.2 \lesssim Q \lesssim 0.8$. 
For example, HW\,20 shows a contamination level of about 24-30\% (see Fig.~\ref{fig:delta_N_memb}) and an increase in $Q$ of about 0.08. 
This $Q$ difference between 0\% and 50\% contamination level in Fig.~\ref{fig:conta} only happens for non-fractal clusters with $\alpha \approx 1$ and fractal clusters with dimension around the limit of $\approx 3$. Therefore, HW\,20 should be non-fractal. Another example is Lindsay\,100 that shows a decrease in $Q$ when the high-contamination is assessed (see Fig.~\ref{fig:delta_Q_memb}, ~\ref{fig:delta_N_memb}). The models of Fig.~\ref{fig:conta} show that the only case where $Q$ decreases with an increasing amount of contamination is for most concentrated clusters with higher $\alpha$ values, and this is exactly the case of Lindsay\,100, which is therefore a non-fractal cluster.
 
We could conclude that if we have fractal clusters contaminated with background stars we are going to measure a $Q$ parameter in the range $Q>0.8$ so not in the fractal region.  
On the other hand, a non-fractal cluster would not become a fractal one by contamination unless the final quantity left of stars is less than 20 as it was shown by \citealt{2019MNRAS.490.2521H}. This is not the case of our sample, as discussed in the previous Section 3.1. Moreover, as we discussed above, the increment in $Q$ measured in our control sample are all smaller than 0.1, which is expected for non-fractal clusters (see Fig.~\ref{fig:conta}), and not for fractal clusters that should have their $Q$ increasing by 0.1-0.4. If we extrapolate these conclusions for the entire sample, we infer that our results will not hide fractal clusters when we use all stars within the cluster radius without field star decontamination.

The next question is which cluster radius we assume to have the purest cluster sample, when we do not have the membership probability information available. To answer this point, we also studied the differences in the estimated $Q$ using the adopted clusters radius ($r_h$, see Sect. \ref{sec:results}) and using $2r_{c}$, $2.5r_{c}$, $3r_{c}$ and $r_{t}$, where $r_c$ is the core radius and $r_{t}$ is the tidal radius determined by \cite{2020MNRAS.498..205S}. In Fig.~\ref{fig:delta_Q_rad} we present the histograms of these differences relative to $r_h$ ($\Delta Q$). We note that for most clusters $\Delta Q<0.1$. The biggest differences happen when we take an excessively large radius like $r_{t}$ or $3r_{c}$. This small variation reinforces the reliability of the results obtained in the present work. 

\section{Results}
\label{sec:results}

\begin{table*}
    \centering
    \caption{Cluster parameters derived in this work. The columns refer to cluster name, equatorial coordinates, number of cluster stars, cluster size R (either $r_h$ or $2.5\times r_c$, see text for details), MST parameters $\overline{m},\overline{s},Q$, chronological age, crossing time and masses (from \citet{2020MNRAS.498..205S}).}
    \label{tab:results}
    \begin{tabular}{c|cccccccccc}
    \noalign{\smallskip}
    \hline\hline
    \noalign{\smallskip}
Object &  RA         & DEC          & N*  &  R         &  $\overline{m}$              &     $\overline{s}$           &   $Q$            & log($T$/yr)  & $T_{c}$  & log(M/M$_{\odot}$) \\
       & (hh mm ss)  & (dd mm ss)   &     & (arcsec)   &                 &                 &                 &              & (Myr)    \\
    \noalign{\smallskip}
    \hline
    \noalign{\smallskip}
SL36   & 04 46 09    & $-$74 53 18  & 101 & 12.50       & 0.648$\pm$0.014 & 0.728$\pm$0.023 & 0.890$\pm$0.026 & 9.30         & 15.83  & 3.38   \\
SL61   & 04 50 45    & $-$75 32 00  & 448 & 38.33    & 0.656$\pm$0.014 & 0.775$\pm$0.023 & 0.846$\pm$0.026 & 9.26         & 36.69   & 4.11 \\
SL576  & 05 33 13    & $-$74 22 08  & 241 & 27.50       & 0.701$\pm$0.014 & 0.852$\pm$0.023 & 0.823$\pm$0.026 & 8.99         & 17.31  & 4.33  \\
SL835  & 06 04 48    & $-$75 06 09  & 69  & 14.17    & 0.735$\pm$0.014 & 0.853$\pm$0.023 & 0.861$\pm$0.026 & 9.30         & ---  & ---    \\
OHSC3  & 04 56 36    & $-$75 14 29  & 38  & 8.33    & 0.723$\pm$0.014 & 0.848$\pm$0.023 & 0.852$\pm$0.026 & 9.25         &  9.90   & 3.26 \\
... & ...    & ... & ...  & ...    & ... & ... & ... & ...         &  ...  & ...  \\
    \noalign{\smallskip}
    \hline
    \noalign{\smallskip}
    \end{tabular}\\
    Note: The full version of this table will be made available as supplementary material online.
\end{table*}

We estimated the value of $Q$ for all the clusters, taking into account only the brightest stars ($V < 22$ mag,  which correspond to a $\sim$~1.3~$M_{\odot}$ main sequence stars in the Magellanic Clouds) with lowest photometric errors in $V$ and $I$ bands ($e_V<0.1$ and $e_I<0.1$). This was the magnitude range that we used in the tests performed in the previous section to conclude that the $Q$ does not change significantly if we consider or not the membership probability of the stars.
Besides, average photometric completeness is above $\gtrsim 80\%$
in the field area and $\gtrsim 60\%$ in the cluster area for V$\approx 22$ mag, and goes up to 100\% for brighter magnitudes \citep[see e.g. Paper I,][]{2019MNRAS.484.5702M}. Therefore, the magnitude cut in $V < $ 22 mag also minimizes errors due to photometric incompleteness. We adopted $r_h$ values estimated by \cite{2020MNRAS.498..205S} as the cluster radius, when it was available. Otherwise, we adopted $2.5r_{c}$ as the cluster radius, since results in Fig~\ref{fig:delta_Q_rad} point out that values of $Q$ using these two radii are generally quite similar. The obtained values of $Q$, $\overline{m}$ and $\overline{s}$, together with other information of the clusters such as the number of stars and the value of the adopted radius are shown in Table~\ref{tab:results}. There we present only the first 5 rows, the complete table is available online. 

The left panel of Fig.~\ref{fig:histQ} shows the histogram of the obtained values of $Q$ for the LMC and for the SMC clusters, whereas the right panel shows the total histogram. From these figures we can see that most of clusters present $Q$ values greater than 0.8. This means that they present a distribution of stars that is homogeneous or with a central concentration and a radial gradient. Only six of the studied clusters ($\sim 7\%$ of the sample) present clear substructures ($Q<0.78$).  Only three of them (SL~28, L~32 and LW~141) have more than 20 stars and therefore reliable $Q$ values (see Sect.~\ref{Sec:MST}). The MST of SL~28 is presented in the right panel of Fig.~\ref{fig:MST}. We can notice the substructure distribution of this cluster as several internal subgroups of stars.  

\begin{figure}
	\centering
	\includegraphics[trim=0 50 0 0, width=\columnwidth]{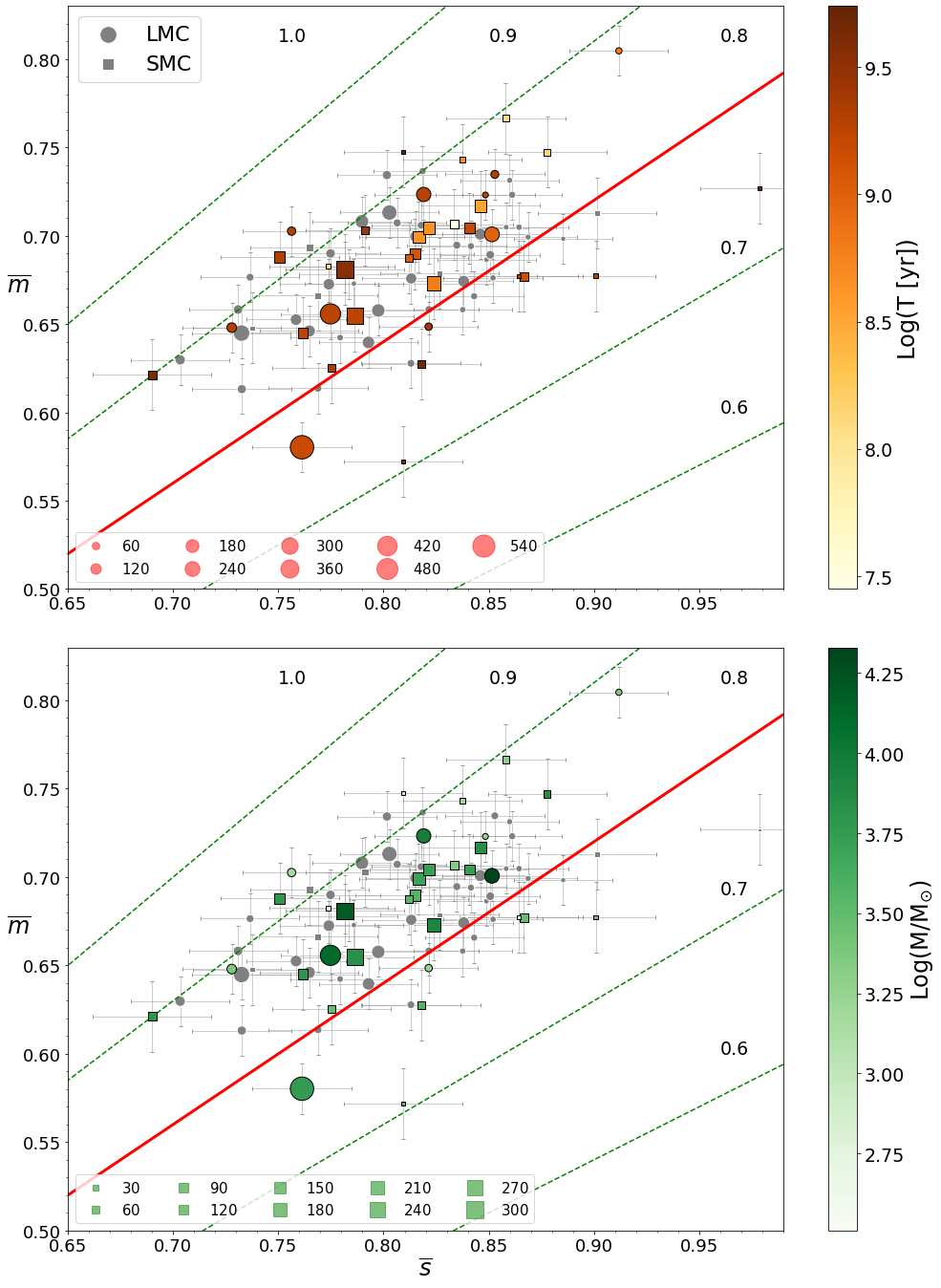}
	\caption{$\overline{m}$ vs. $\overline{s}$ diagram for all the studied clusters, LMC clusters are represented with a circle and SMC clusters with a square. The colour bar in the upper panel indicates ages, and in the bottom panel masses whenever those quantities available from \citet{2020MNRAS.498..205S}. Clusters without measured age or mass are indicated in gray. Errors were computed using the bootstrap method. The solid red line indicates $Q=0.8$, and the dashed green lines indicate other values of $Q$. Symbol size indicates the number of stars in the cluster, being 14 the minimum number of stars in a cluster and 599 the maximum. Some examples of the size corresponding to a certain number of stars are shown in the labels. }
	\label{fig:m_s}
\end{figure}

The $\overline{m}$ vs. $\overline{s}$ diagram is a useful tool in the study of the internal structure of clusters. Studies with simulated clusters \citep[e.g.][]{2004MNRAS.348..589C, 2020MNRAS.493.4925D} reveal the location in the $\overline{m}$ vs. $\overline{s}$ diagram of each cluster population (radial or fractal). We present in Fig.~\ref{fig:m_s} the $\overline{m}$ vs. $\overline{s}$ plot for the clusters studied in this work. 
The clusters are mostly distributed between the iso-$Q$ lines of 0.8 and 0.9, as it was shown in Fig. \ref{fig:histQ}, but now revealing other parameters. Most of the studied clusters follow the theoretical behavior predicted for clusters with radial distribution of stars \citep[see e.g. Fig.~1 in][]{2009MNRAS.400.1427C}.
Cluster ages are available for sub-sample of the clusters from \citet{2020MNRAS.498..205S}, and they indicate a mild trend in the sense that there seem to be older clusters toward larger (smaller) values of $Q$ ($\overline{s}$),
which corresponds to more centrally concentrated cluster models according to \citet{2009MNRAS.400.1427C}. We 
will discuss this apparent trend with cluster age in more detail in Sect.~\ref{sec:disc_a}.

\begin{figure*}
	\centering
	\includegraphics[width=0.9\textwidth]{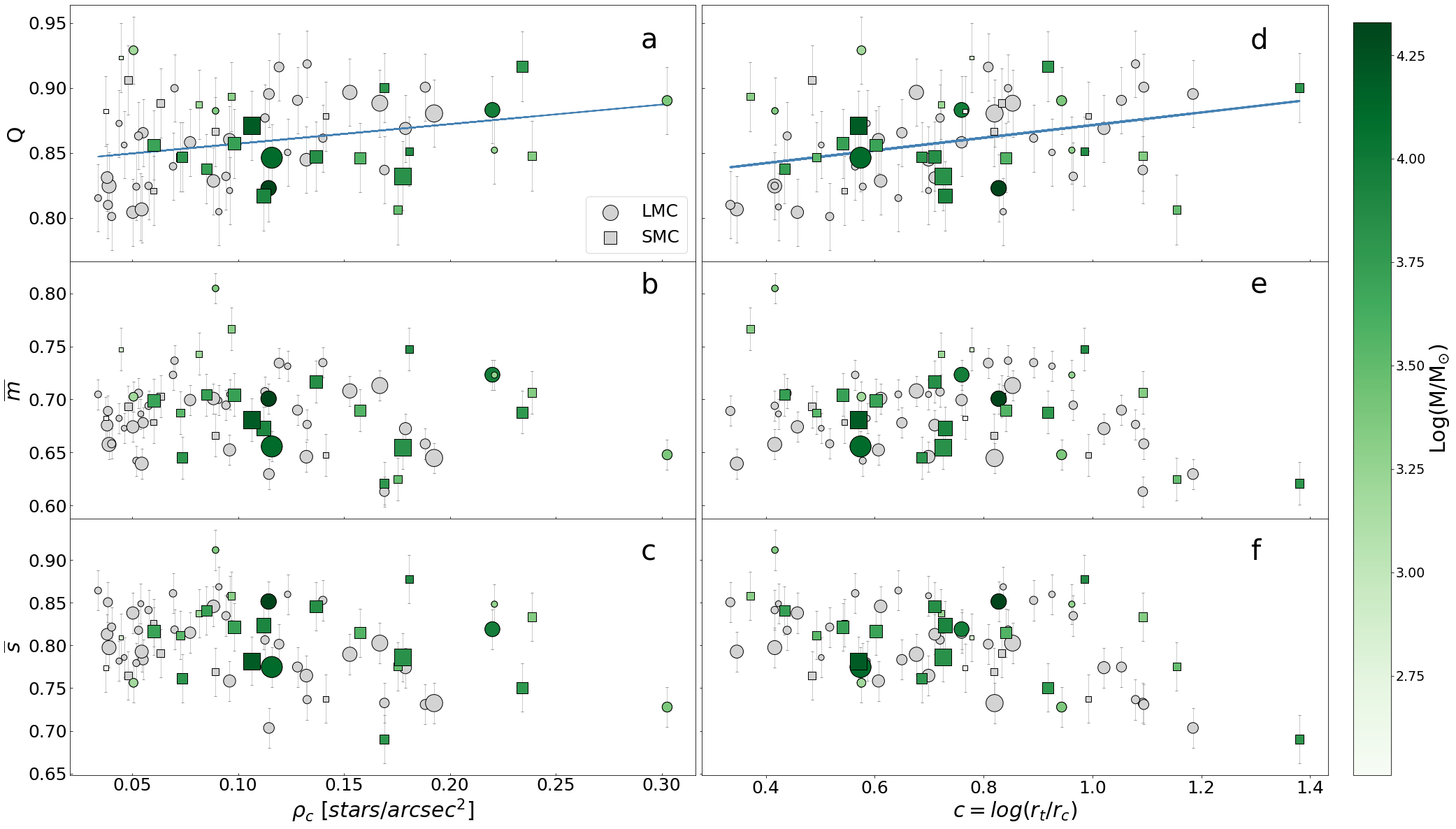}
	\caption{Variation of  $Q$, $\overline{m}$ and $\overline{s}$ with the stellar density in the cluster core ($\rho_c$, panels a, b and c respectively), and with the concentration parameter ($\log{(r_t/r_c)}$, panels d, e and f respectively).  As in Fig.~\ref{fig:m_s}, the sizes of the symbols indicate the number of stars in each cluster. Circles and squares represent the LMC and SMC clusters respectively. The color bar indicate masses.
	The blue line in panels a and d corresponds to a linear fit and points a positive trend between the two quantities. }
	\label{fig:parameters_rc}
\end{figure*}

The simulations by \citet{2009MNRAS.400.1427C} also show some trends of cluster concentration with $Q, \overline{m}, \overline{s}$.
For those clusters with a central concentration of stars ($Q>0.8$), we studied the variation of the parameters $Q, \overline{m}$ and $\overline{s}$ with the surface stellar density of the core $\rho_c$, which was estimated counting the number of stars inside $r_c$ (panels a, b and c of Fig.~\ref{fig:parameters_rc}), and with the concentration parameter $c=\log(r_t/r_c)$ (panels d, e and f of Fig.~\ref{fig:parameters_rc}). In general, the distributions have a large dispersion, but some trends are clearly visible. The value of $Q$ seems to increase with $\rho_c$, i.e.,
clusters with the densest core tend to possess higher values of $Q$. Additionally, $Q$ also seems to increase with the concentration parameter $c$. $Q$ would be higher in clusters with well defined radial density profiles. 
On the other hand, $\overline{m}$ and $\overline{s}$  seem to present a slight negative trend in which these values decrease while the core density and concentration parameter increase.
Finally, the trends with $c$ seem to be better defined than those with $\rho_c$. This could be an artifact caused by a relatively small number of clusters, but if confirmed, this conclusion would mean that the relative radial distribution 
of stars is more important than the absolute central density of stars when the goal is to characterize
non-fractal clusters with MST.

\begin{figure*}
	\centering
	\includegraphics[width=0.9\textwidth]{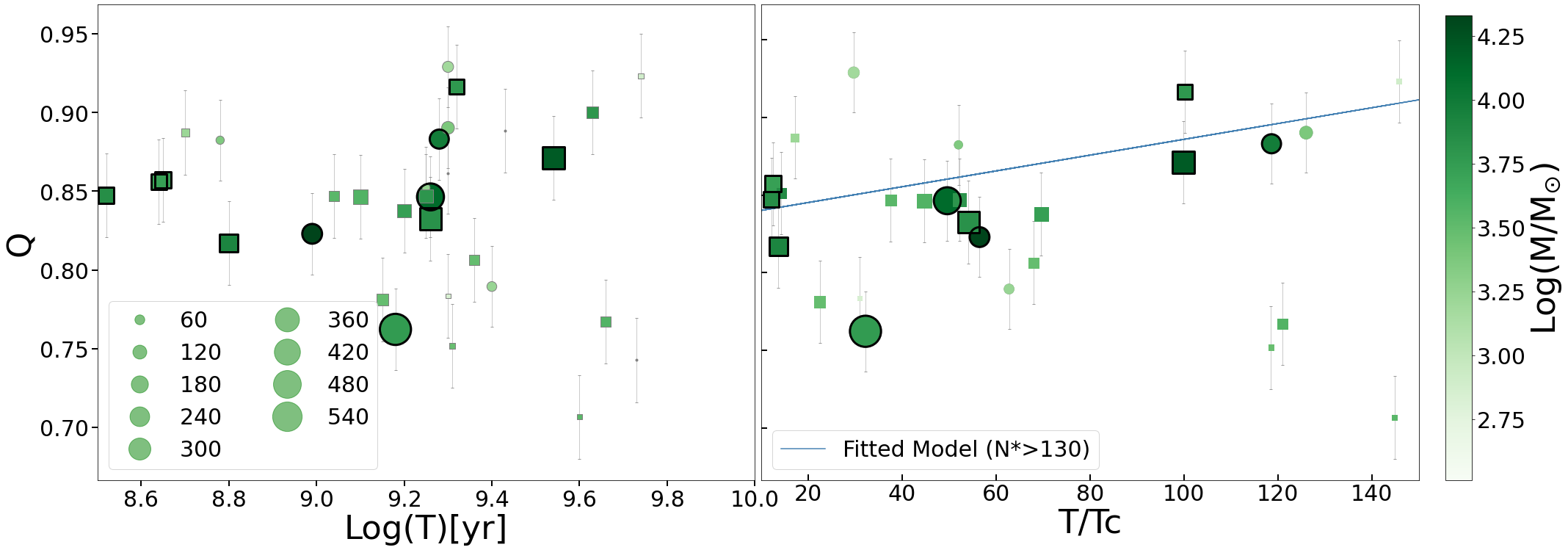}
	\caption{\emph{Left:} $Q~vs.~Log(T)$ for all the studied clusters with measured ages. As in previous figures circles and squares represent clusters in the LMC and SMC respectively. The clusters with black border have more than 130 stars. The sizes of the symbols represent the number of stars in the clusters  (examples are shown in the label).
	\emph{Right:} $Q~vs.~T/Tc$. 
	The blue line indicates a linear fit over the most populated clusters (symbols with black border). This trend indicates that clusters more dynamically evolve have larger values of $Q$. Clusters were colored according to their masses.}
	\label{fig:qage}
\end{figure*}

\section{Discussion}
\label{sec:discussion}

\subsection{Dynamical age}
\label{sec:disc_a}

Taking into account theoretical models in which a set of particles with a given mass distribution and dynamically evolve from a fractal structure, these models develop paths in a $\overline{m}$ vs. $\overline{s}$ diagram, crossing the limit $Q = 0.8$ and ending with a spatial distribution showing a central concentration. However, the time it takes for such models to carry out that path strongly depends on the initial ratio between kinetic and potential energies \citep{2012MNRAS.427..637P}. This relation was explored by several authors by means of both numerical and observational works. Among them, the works of \cite{2009ApJ...696.2086S} and \cite{2019MNRAS.490.2521H} carried out on clusters of the Milky Way. In particular, \cite{2009ApJ...696.2086S} studied the internal structure of 16 open clusters in the Milky Way with ages between 8 Myr and 4 Gyr.
They found hints of a correlation in which the $Q$ parameter grows with the dynamical age of the clusters. Additionally, \cite{2008MNRAS.391L..93G} and \cite{2009MNRAS.392..868B} carried out similar studies over different populations in the MC. We have shown a mild trend of chronological age with $Q$ in Fig. \ref{fig:m_s}, and we now analyze whether the dynamical age reveals similar trends with respect to the aforementioned works.


The left panel of Fig.~\ref{fig:qage} shows $Q~vs.~log(T)$, where $T$ is the chronological age in years. From this figure, it is not possible to see a clear relationship between these two quantities. 
However, the dynamical evolution of a stellar cluster is a complex process that involves several factors like the total kinetic energy of the system, the potential gravitational energy of the total mass, two-body interactions, the amount of gas that is
removed from the system in the initial stage and the rate in which it is removed, the galactic environment in which they are located, etc \citep[e.g.][]{2009ApJ...704..453F, 2019ARA&A..57..227K}. 
Therefore, not all clusters are in the same stage of dynamical evolution in the same period of time since their formation.
For this reason, we should study the behavior of $Q$ with the dynamical age of the cluster.
\citet{2011MNRAS.410L...6G} used the cluster age over its crossing time ($T_{c}$) as an estimator of the dynamical age of the cluster.
This quantity gives how many crossing times the cluster has undergone,
which indicates how much the stars have interacted since their formation. 
We used the expression given in \cite{2011MNRAS.410L...6G} as a good approximation for $T_{c}$:

$\displaystyle T_{c}=10~\Bigg(\frac{r_{h}^3}{G~M}\Bigg)^{1/2},$

\noindent where $G$ is the gravitational constant and $M$ is the total mass of the cluster. 
\cite{2011MNRAS.410L...6G} use the dynamical age to separate bound clusters from unbound stellar association. Stellar groups with ages exceeding the crossing time are bound. Otherwise if the age of the cluster is lower than its crossing time we have unbound associations.
 From the right panel of Fig.~\ref{fig:qage} we can see that the sample of systems with measured ages studied here have ages that exceeds several times the value of their crossing time, i.e., $T/T_{c} >$1. Therefore, all the clusters studied in the present work are bound.

The right panel of Fig.~\ref{fig:qage} shows the variation of $Q$ with the dynamical age. There seem to be a positive correlation between these two quantities, which is more evident if we see only the cluster with more than 130 stars (points with black border in the figure), which are expected to have more reliable $Q$ values (see Sect~\ref{contamination}). We were able to fit a linear model (blue line) to the data taking into account only these more populous clusters.  
This result indicates that more dynamically evolved clusters, i.e. those clusters that have lived more
crossing times, appear to have higher values of $Q$. Values of ages and ${T_c}$ for the clusters are listed in Table~\ref{tab:results}.

 \begin{figure}
	\centering
	\includegraphics[width=1.0\columnwidth]{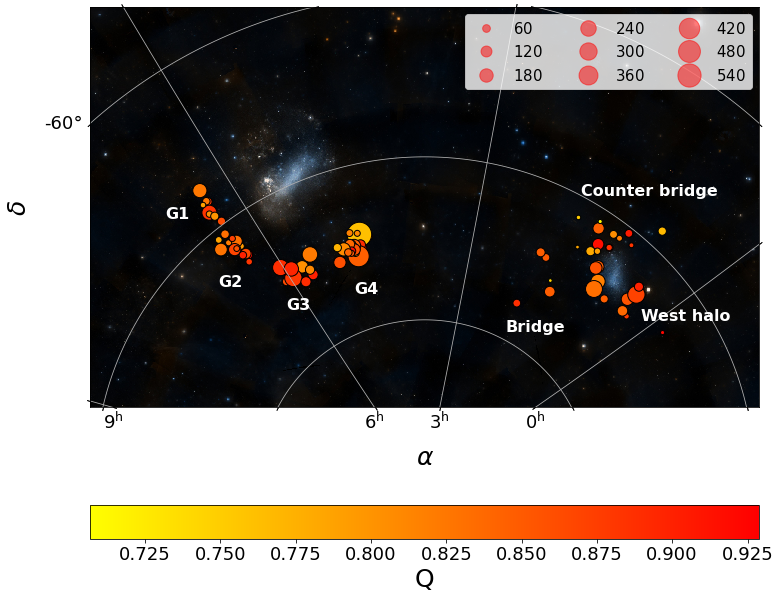}
	\caption{Spatial distribution of clusters overlapped in a DSS2 color image. The different groups of clusters are indicated according to \citet{2014A&A...561A.106D, 2020MNRAS.498..205S}. The colour bar indicates the Q values. The size of the symbols indicates the  number of stars in the clusters as in Fig.~\ref{fig:m_s}.}	
	\label{fig:Sdist}
\end{figure}

\subsection{Spatial distribution}
\label{sec:disc_b}

Fig.~\ref{fig:Sdist} shows the cluster distribution on the sky, where the value of $Q$ is indicated with the colour bar. The different groups of clusters are indicated according to \citet{2014A&A...561A.106D,2020MNRAS.498..205S} for the SMC and LMC, respectively.
The most populous clusters are in the regions G3 and G4 of the LMC and in the West halo and inner part of the Bridge for the SMC. So, in these regions the obtained values of $Q$, $\overline{m}$ and $\overline{s}$ are more reliable.

\begin{figure*}[h]
	\centering
	\includegraphics[width=0.9\textwidth]{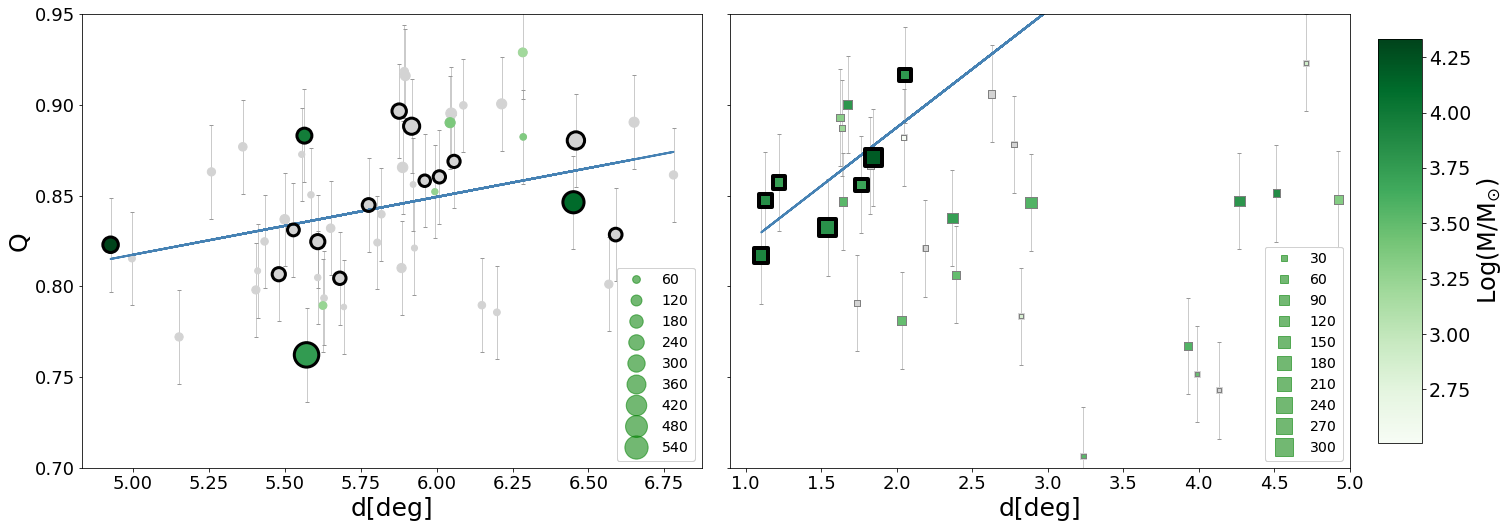}
	\caption{Q vs. projected galactocentric angular distance for clusters in the LMC (\emph{Left}) and SMC (\emph{Right}). Symbol size indicates the number of stars in the clusters (examples are shown in the labels). A linear fit to the clusters with more than 130 stars (black border) is shown. Color bar indicates the cluster masses.}
	\label{fig:Qdist}
\end{figure*}

Figure~\ref{fig:Sdist} also shows a mild
correlation in which clusters further away from the center of the galaxy seem to have higher 
values of $Q$. It means that they would present a better defined central concentration of stars with a radial gradient. 
To properly 
investigate this relation, we plot $Q$ against the projected angular galactocentric distance for the LMC (left) and SMC (right) in Fig.~\ref{fig:Qdist}.
In the left panel we can observe a slight general trend in which $Q$ increasebecause the line-of-sight depth of the SMC is much larger than its projected size on sky \citep{Crowl2001AJ....122..220C, Glatt2008AJ....136.1703G, Subramanian2012ApJ...744..128S, 2022MNRAS.512.4334D}. Cluster distances derived within the VISCACHA survey will be required to check this relation with 3D distances. 
Taking into account only the clusters with more than 130 stars (points surrounded by a black circle in the right panel of Fig.~\ref{fig:Qdist}), we were able to fit a linear relationship between these two quantities even for the SMC (blue lines). 
As larger values of $Q$ correspond to more dynamically evolved clusters, this could be indicating that the most distant clusters from the center of each galaxy are more evolved in a dynamical sense, at least within the galaxy tidal radius.

A dependence of $Q$ with the galactocentric distance is expected since the gravitational tidal field of the galaxy on a star in the cluster is weaker at  
larger galactocentric distance. This directly influencesinfluencers the rate at which stars escape from the system \citep{Madrid_2012} and therefore its internal structure and dynamical evolution.
Clusters nearer the galactic centre, if they survive, should have undergone stronger mass loss, which decreases their binding energy causing overall expansion (leading to a more uniform stellar distribution where $Q \sim$ 0.8).
Consequently, in the less dense internal environment, this process delays the cluster dynamical evolution as driven by two-body relaxation.

In the LMC, all the studied clusters are inside the tidal radius. So it is possible to observe the trend mentioned above with some dispersion due to the narrow range of galactocentric distances. It will be necessary to study clusters spread over a wider range of distances in order to better investigate this relationship. 
The tidal radius of the galaxy for the SMC is around $d\approx3-4^{\circ}$ \citep{2021A&A...647L...9D}. Clusters within this radius show similar behavior to that observed in the LMC, i.e. we observe a large dispersion but it is possible to observe a trend for the most populous clusters. The clusters beyond this radius show a behavior that is a mix of projection effect and tidal forces which have a strong influence in the SMC \citep{2022MNRAS.512.4334D}.

\section{Summary and conclusions}
\label{sec:conclusions}

We have derived the parameters $\overline{m}$, $\overline{s}$ and $Q$ by means of the MST method, which are related to
the internal structure of star clusters. These parameters are
useful for distinguishing between fractal and centered distributions of stars. We derived these quantities for the 82 clusters in the VISCACHA  Survey covering the LMC and SMC outer regions. 
We found that most of the studied clusters present a homogeneous distribution of stars ($Q\sim0.8$) or a distribution with a central concentration ($Q>0.8$). Only 3 clusters from the sample present a fractal distribution of stars ($Q<0.8$). For 5~\% of the sample we were not able to obtain reliable values due to the low number of stars.

We obtained that the value of $Q$ seems to grow with the superficial density of stars in the core and also with the concentration parameter. Additionally, we found that the location of the clusters in the $\overline{m}-\overline{s}$ diagram is in agreement with those obtained for simulated clusters with a central concentration of stars. On the other hand, analyzing only the clusters with more stars ( with more reliable
$Q$ values), 
we found a correlation between $Q$ and the galactocentric distance for each galaxy within their tidal radii.

The studied sample covers a wide range of ages from 40~Myr to 5.5~Gyr which allowed us to study the evolution of the internal structure of a cluster with age. Our results indicate that populous clusters more evolved dynamically tend to present large values of $Q$. There were only three clusters showing possible fractal structure with $Q < 0.8$. Therefore, the present sample did not allow us to set a threshold in age beyond which a cluster loses its initial fractal structure.

The present study provides several constraints to better understand the dynamical evolution of star clusters in general and the structure of the outer regions of the MCs in particular. Nevertheless, the time when a cluster loses the fractal information was not yet fully characterized with observations, and we propose that a larger sample including a larger number of younger and less evolved clusters, as well as covering a larger range of distances from the LMC center should be analyzed in a follow-up work.

\section*{Acknowledgements}

This research has received financial support from the Argentinian institutions
UNLP (Programas de Incentivos 11/G158 and 11/G168); CONICET (PIP 112-201701-00015, PIP 112-201701-00055),SECYT (Universidad Nacional de Córdoba) and Agencia Nacional de Promoción Científica y Tecnológica (ANPCyT) I+D+i (PICT 2019-0344). 
This study was financed in part by the Coordena\c c\~ao de Aperfei\c coamento de Pessoal de N\'ivel Superior - Brasil (CAPES) - Finance Code 001. JFCJS acknowledges support by FAPEMIG (proc. OET-00020-22).
B.D. acknowledges support by ANID-FONDECYT iniciación grant No. 11221366. 
F. Maia acknowledges financial support from Conselho Nacional de Desenvolvimento Científico e Tecnológico - CNPq (proc. 404482/2021-0) and from FAPERJ (proc. E-26/201.386/2022 and E-26/211.475/2021).
D.M. gratefully acknowledges support by the ANID BASAL projects ACE210002 and FB210003 and by Fondecyt Project No. 1220724.
R.A.P.O. acknowledges the FAPESP PhD fellowship no. 2018/22181-0.
A.P.-V. acknowledges the DGAPA-PAPIIT grant IA103122.
Based on observations obtained at the Southern Astrophysical Research (SOAR) telescope, which is a joint project of the Minist\'erio da Ci\^encia, Tecnologia, e Inovação (MCTI) da Rep\'ublica Federativa do Brasil, the U.S. National Optical Astronomy Observatory (NOAO), the University of North Carolina at Chapel Hill (UNC), and Michigan State University (MSU).
The authors want to thank for the use of the NASA Astrophysics Data System.


\section*{Data availability}

The data underlying this article are available in the NOIRLab Astro Data Archive (https://astroarchive.noirlab.edu/) or upon reasonable request to the authors.



\bibliographystyle{mnras}
\bibliography{biblio} 







\bsp	
\label{lastpage}
\end{document}